\documentclass[11pt]{article}
\usepackage{moriond,epsfig}

\newcommand{\epem}              {\ensuremath{\mathrm{e^+e^-}}}
\newcommand{\roots}             {\ensuremath{\sqrt{s}}}
\newcommand{\as}                {\ensuremath{\alpha_\mathrm{S}}}
\newcommand{\mz}                {\ensuremath{M_{\mathrm{Z^0}}}}
\newcommand{\asmz}              {\ensuremath{\as(\mz)}}

\newcommand{\bbbar}             {\ensuremath{\mathrm{b\bar{b}}}}

\newcommand{\stat}              {\ensuremath{\mathrm{(stat.)}}}

\newcommand{\expt}              {\ensuremath{\mathrm{(exp.)}}}
\newcommand{\had}               {\ensuremath{\mathrm{(had.)}}}
\newcommand{\theo}              {\ensuremath{\mathrm{(theo.)}}}
\newcommand{\thr}               {\ensuremath{1-T}}
\newcommand{\mh}                {\ensuremath{M_\mathrm{H}}}
\newcommand{\bt}                {\ensuremath{B_\mathrm{T}}}
\newcommand{\bw}                {\ensuremath{B_\mathrm{W}}}
\newcommand{\cp}                {\ensuremath{C}}
\newcommand{\ytwothree}         {\ensuremath{y_{23}}}

\newcommand{\chisq}  {\ensuremath{\chi^2}}
\newcommand{\chisqd} {\ensuremath{\chi^2/\mathrm{d.o.f.}}}
\newcommand{\xmu}      {\ensuremath{x_{\mu}}}

\begin{document}
\vspace*{4cm}
\title{{\boldmath\asmz} FROM JADE EVENT SHAPES}

\author{S. Kluth}

\address{Max-Planck-Institut f\"ur Physik, D-80805 M\"unchen, Germany,
skluth@mpp.mpg.de}

\maketitle\abstracts{Event shape data from \epem\ annihilation 
into hadrons collected by the JADE experiment between $\roots=14$ and
44~GeV are used to determine the strong coupling \as.  QCD predictions
complete to next-to-next-to-leading order (NLLO), alternatively
combined with next-to-leading-log-approximation (NLLA) are used.  The
stability of the NNLO and NNLO+NLLA results with respect to variations
of the renormalisation scale is improved compared to previous results
obtained with next-to-leading-order (NLO) or NLO+NLLA predictions.
The energy dependence of \as\ agrees with the QCD prediction of
asymptotic freedom and excludes absence of running with 99\%
confidence level.}

\section{Introduction}

The determination of the strong coupling \as\ in Quantumchromo
Dynamics (QCD) is an important test of the
theory~\cite{bethke06,kluth06,pdg08}.  The analysis of hadron
production in \epem\ annihilation with observables based on jet
production and event shape definitions plays a major role, because in
the \epem\ environment there is no interference between intial and
final state and one generally has a clean experimental environment
with small background rates and neglegible pileup.  Recently the NNLO
corrections for QCD predictions of event shape observable
distributions~\cite{gehrmannderidder07b,weinzierl08} and
moments~\cite{gehrmannderidder09} were completed and used for a
measurement of \as\ using data of the LEP experiment
ALEPH~\cite{dissertori07}.  It is therefore of high interest to
measure \as\ with JADE data and the same improved QCD
predictions~\cite{jadennlo}.  We also include matching of existing
NLLA QCD calculations with the new NNLO
calculations~\cite{gehrmann08}.

The JADE experiment operated at the PETRA \epem\ collider at DESY from
1979 to 1986 at centre-of-mass (cms) energies $\roots=12$ to 44~GeV.
Significant data samples were recorded at $\roots=14, 22, 34.6, 35,
38.3$ and 43.8~GeV.  The JADE detector is described
in~\cite{naroska87}.  The JADE experiment combines tracking of charged
particles in a solenoidal magnetic field and electromagnetic
calorimeters over a large solid angle.  This allows to select hadronic
final states with high efficiency while demanding that the events are
fully contained in the detector.  The JADE reconstruction, simulation
and analysis software was ported to current computers~\cite{pedrophd}
and is used in our work.

At the JADE energies background processes relevant for hadron
production are $\epem\rightarrow\tau^+\tau^-$ with subsequent hadronic
$\tau$ decays and two-photon interactions with hadronic final states.
Initial state radiation (ISR) of photons from the beam particles also
occurs with a significant rate leading the production of hadronic
final states at reduced cms energy.  After application of suitable
event selection cuts based on 4-momentum balance and particle
multiplicity the background processes can be neglegted.  Residual
effects of ISR are corrected for.

We use the event shape observables thrust \thr, heavy jet mass \mh,
total and wide broadening \bt\ and \bw, C-parameter \cp\ and
\ytwothree, the value of the distance definition
$y_{ij}=2\min(E_i^2,E_j^2)(1-\cos\theta_{ij})/s$ of the Durham jet
algorithm where the event changes from a three- to a two-jet
configuration~\cite{dasgupta03,jadennlo}.

\section{Data Analysis}

The distributions of event shape observables are computed from all
selected events, and from samples of simulated events at several
levels.  The final state after termination of the parton shower is
called {\em parton-level}, the final state after hadronisation and
decays of particles with lifetimes shorter than 300~ps is called {\em
hadron-level}, and after passing the events through the JADE detector
simulation, recontruction and analysis procedures the result is called
{\em detector-level}.  We use the programs PYTHIA, HERWIG and ARIADNE
tuned by the OPAL collaboration for this purpose, see~\cite{jadennlo}
for details.

The contribution of $\epem\rightarrow\bbbar$ events is subtracted from
the distributions using simulated events at detector-level.  The
results are multiplied by the ratio of simulated hadron- and
detector-level distributions to correct for experimental acceptance
and resolution.  In this way we obtain distributions of event shape
observables corrected to the hadron-level.  A comparison of observed
and corrected event shape distributions with predictions by the Monte
Carlo simulations shows satisfactory agreement.

\section{Determination of \as}

\begin{figure}[htb!]
\begin{tabular}{cc}
\includegraphics[width=0.45\columnwidth]{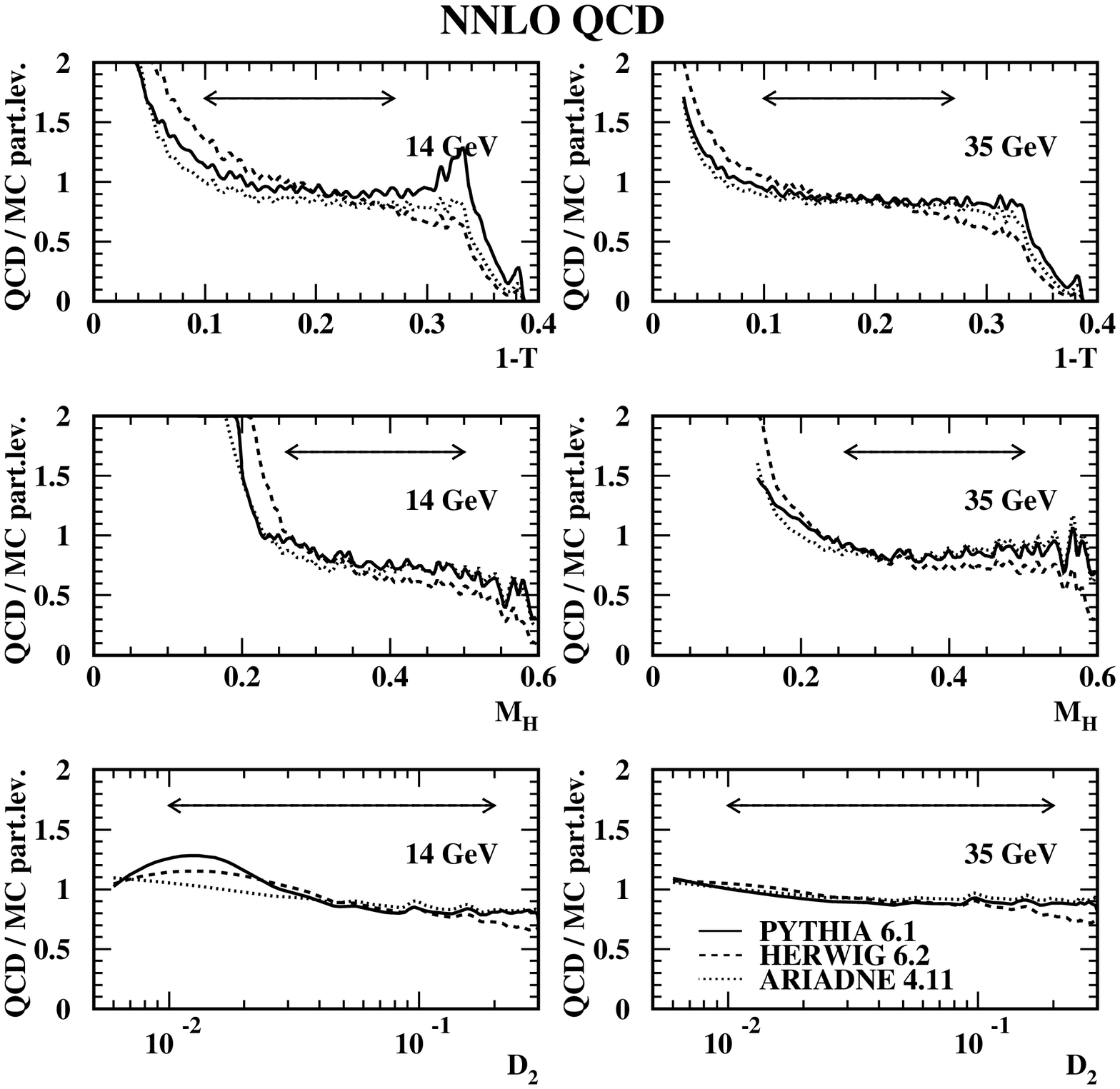} &
\includegraphics[width=0.45\columnwidth]{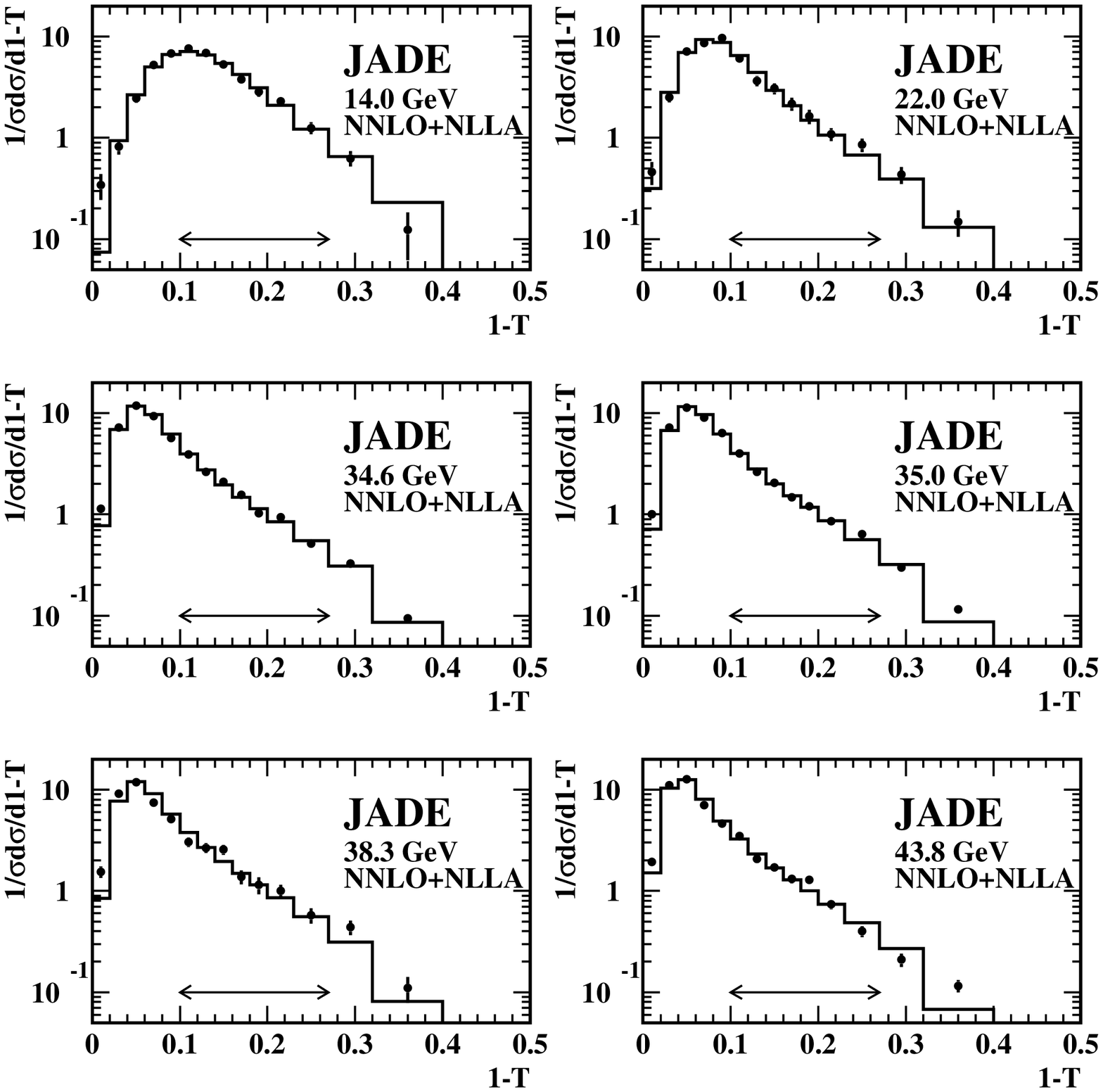} \\
\end{tabular}
\caption[bla]{(left) Ratio of QCD and MC prediction at parton-level
for observables and cms energies as indicated on the figures.  The
different line types show the ratios for different MC programs.
(right) Data at hadron-level for \thr\ with superimposed NNLO+NLLA
fits at the six JADE energy points. The arrows correspond to the fit
ranges.}
\label{fig_mcqcd}
\end{figure}

The hadron-level distributions are compared with the QCD predictions
corrected for hadronisation effects by a \chisq-method with \as\ as a
free parameter and the full statistical covariance matrix.  The
hadronisation corrections for the QCD predictions are found for
cumulative theory predictions by dividing cumulative distributions at
hadron- and parton-level.  In order to justify this procedure we
compare simulated distributions at parton-level with the NNLO QCD
calculations for $\asmz=0.118$ and find reasonable agreement as shown
in figure~\ref{fig_mcqcd} (left).  The program PYTHIA is used in the
standard analysis and HERWIG and ARIADNE are used as alternatives.

Only restricted regions of the distributions are used for the fits,
i.e.\ those where the experimental and hadronisation corrections are
stable and where the QCD predictions have LO contributions and don't
have large logarithmically enhanced terms~\cite{jadennlo}.  The resulting
fit ranges are the same for all energy points and are shown on
figure~\ref{fig_mcqcd} (right) for \thr.

The experimental uncertainties are found by repeating the analysis
with varied selection cuts and procedures.  The uncertainties
from the hadronisation corrections are determined by comparing
fit results obtained with PYTHIA, HERWIG or ARIADNE.  The theoretical
uncertainties are found by changing the renormalisation scale
parameter $\xmu=\mu/\sqrt{s}$ from its standard value $\xmu=1$
to $\xmu=0.5$ or 2.0.

Figure~\ref{fig_mcqcd} (right) shows the NNLO+NLLA QCD predictions
using the fit results for \as\ superimposed over the hadron-level data
for \thr.  The agreement between data and theory is excellent for both
NNLO and NNLO+NLLA fits.  For example, the \chisqd\ values for \thr\
with NNLO fits range from 0.7 (14~GeV) to 2.5 (34.6~GeV).  The
root-mean-square (RMS) values of \as\ results from NNLO+NLLA fits to
the six observables are between 0.007 (22~GeV) and 0.003 (44~GeV),
consistent with the theory uncertainties of the combined values (see
below).  This is an important cross check made possible by having
six different observables in the analysis.

\begin{figure}[htb!]
\begin{tabular}{cc}
\includegraphics[width=0.45\columnwidth]{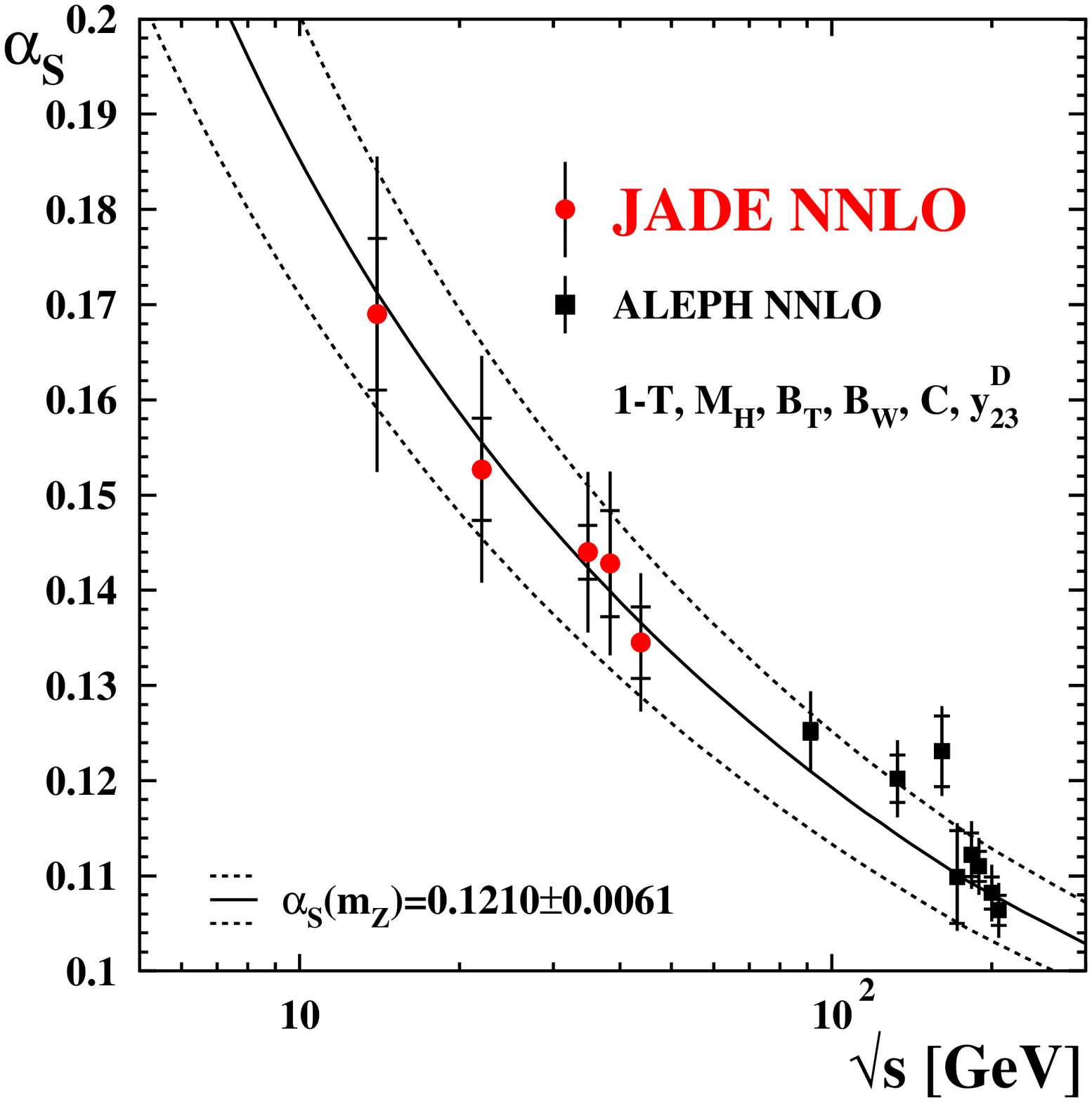} &
\includegraphics[width=0.45\columnwidth]{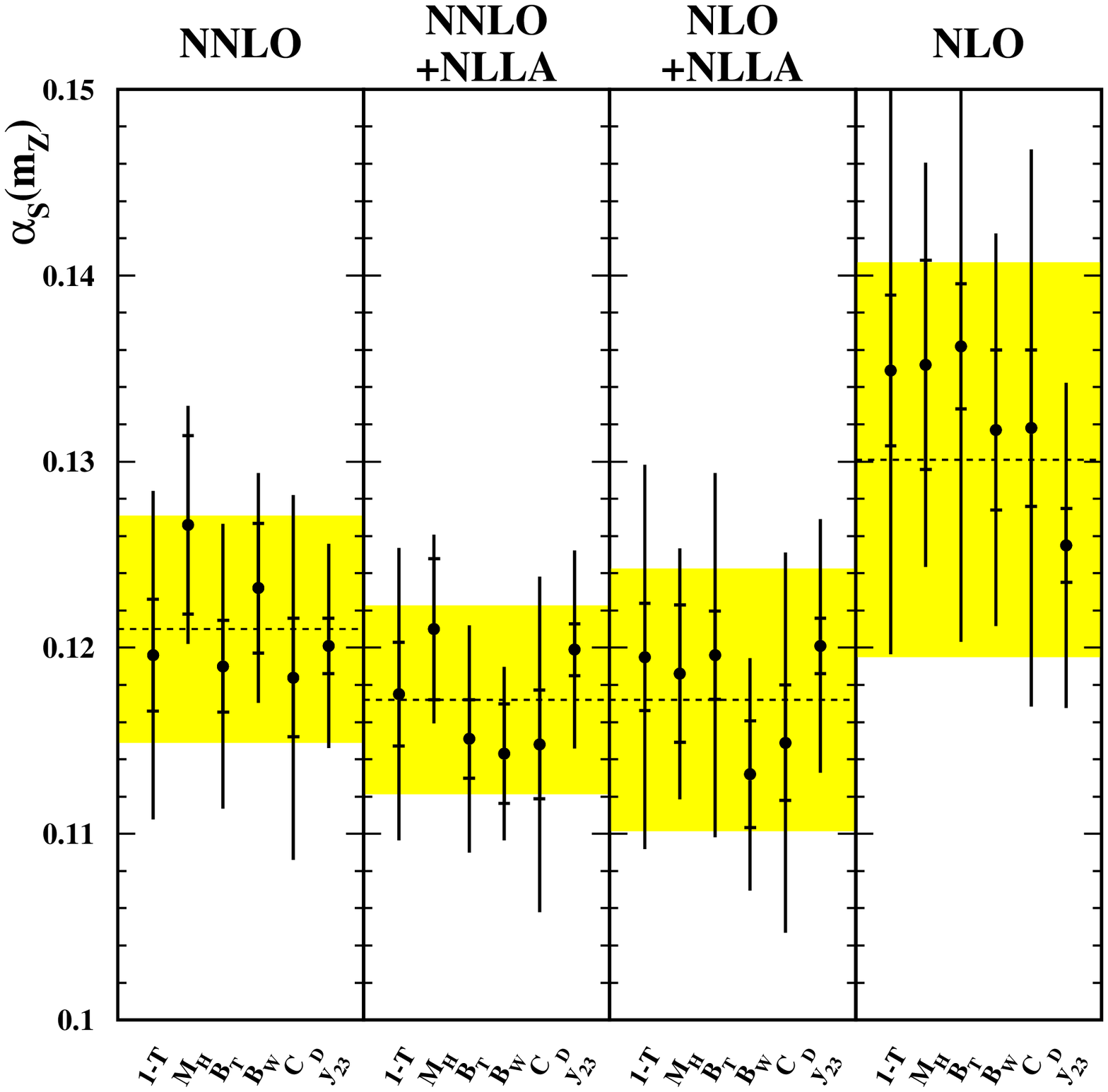} \\
\end{tabular}
\caption[bla]{(left) Combined values of \as\ from NNLO fits
at the JADE energy points together with the results
of~\cite{dissertori07}.  The lines show the evolution of \as\ based on
the combined result.  (right) Combined values of \asmz\ for each
analysis. The shaded bands and dashed lines show the combined values
of \asmz\ with total uncertainties.}
\label{fig_asplot}
\end{figure}

We combine the results from the six observables at each energy point
as in~\cite{OPALPR404,kluth06} and show the results of the NNLO
analysis in figure~\ref{fig_asplot} together with the results
of~\cite{dissertori07}.  The result of combining the combined
values of \as\ after evolving to the reference scale \mz\ is 
shown as lines in the same figure.  The final result
using this combination is for NNLO
\begin{displaymath}
\asmz= 0.1210\pm0.0007\stat\pm0.0021\expt\pm0.0044\had\pm0.0036\theo
\end{displaymath}
and for NNLO+NLLA
\begin{displaymath}
\asmz= 0.1172\pm0.0006\stat\pm0.0020\expt\pm0.0035\had\pm0.0030\theo\;\;.
\end{displaymath}

Our final result from NNLO+NLLA predictions has the smaller
uncertainties of 4\%.  It is consistent with previous determinations
of \asmz\ using NNLO theory~\cite{dissertori07} and with with recent
averages of \asmz~\cite{bethke06,kluth06,pdg08}.  Repeating the
combination of the combined NNLO+NLLA results at each energy point
without evolving to a common reference scale results in a \chisq\
probability of $10^{-2}$ assuming partially correlated hadronisation
uncertainties and ignoring theory uncertainties.  We interpret this as
strong evidence for the dependence of the strong coupling on cms
energy as predicted by QCD from JADE data alone.

Comparison with NLO and NLO+NLLA fits using the same data,
hadronisation corrections and fit ranges gives consistent results.
The theory uncertainty of the NLO+NLLA analysis is larger by about
60\% compared with our NNLO or NNLO+NLLA results.  The theory
uncertainty of the NLO analysis with fixed renormalisation scale
$\xmu=1$ is larger by a factor of 2.6 compared with the NNLO analysis.
Figure~\ref{fig_asplot} (right) shows a summary of the different
analyses.  The smaller theory uncertainties of the NNLO or NNLO+NLLA
analyses compared with NLO or NLO+NLLA analyses follow from the weaker
dependence of the NNLO calculations on the renormalisation scale
parameter \xmu~\cite{jadennlo}.

\section{Summary}

We have shown a determination of the strong coupling constant
\asmz\ with NNLO and NNLO+NLLA QCD calculations using JADE data.  The
final result $\asmz=0.1172\pm0.0051$ has a competetive
uncertainty of 4\% and is consistent with other measurements.  Our
data confirm the running of \as\ as predicted by QCD.


\end{document}